\documentclass[
    ,final            % use final for the camera ready runs
  ]
  {aipproc}

\layoutstyle{8x11single}

\usepackage{amsmath,amssymb}

\usepackage{array}
\usepackage{slashed}
\usepackage{multirow}

\newcommand{\numparents}{\ensuremath{N}}
\newcommand{\flox}{{\cal F}}
\newcommand{\genericT}{\ensuremath{T}}
\newcommand{\ourT}{\ensuremath{\top}} 
\newcommand{\ourperp}{\ensuremath{\vee}}
\newcommand{\massless}{{\ensuremath{\circ}}}
\newcommand{\mmass}{{\slashed{\comp M}}}
\newcommand{\mpt}{\slashed{p}_\genericT}
\newcommand{\comp}[1]{\ensuremath{{\mathbf #1}}}

\newcommand{\ourVec}[1]{\left( #1 \right)}

\newcommand{\mptvec}{\slashed{\vec{p}}_\genericT}

\newcommand{\MEFF}{\ensuremath{m_{\mathrm{eff}}}}
\newcommand{\parent}[1]{\ensuremath{{\mathbb{P}_{#1}}}}
\newcommand{\beq}{\begin{equation}}
\newcommand{\eeq}{\end{equation}}
\newcommand{\ourSet}[1]{\left\{ #1 \right\}}

\newcommand{\ourMaxMinBracs}[1]{\left[ #1 \right]}
\newcommand{\bea}{\begin{eqnarray}}
\newcommand{\eea}{\end{eqnarray}}
\newcommand{\chiM}{\ensuremath{\mmass}}
\newcommand{\parentset}{\ensuremath{\mathcal{P}}}
\newcommand{\invisassign}[1]{\ensuremath{{\mathcal{I}_{#1}}}}

\begin{document}

\title{A Storm in a ``T'' Cup}

                
                % 11.80.Cr	Kinematical properties (helicity and invariant amplitudes, kinematic singularities, etc.)
                % 12.60.-i	Models beyond the standard model (for unified field theories, see 12.10.-g)
	       %11.30.Pb	Supersymmetry (see also 12.60.Jv Supersymmetric models)


\author{A.~Barr}{
  address={Department of Physics, Denys Wilkinson Building, Keble Road, Oxford OX1 3RH, UK}
}

\author{T.~Khoo}{
  address={Department of Physics, Cavendish Laboratory, JJ Thomson
Avenue, Cambridge, CB3 0HE, UK}
}

\author{P.~Konar}{
  address={Theoretical Physics Group, Physical Research Laboratory,
%Navrangpura, 
Ahmedabad, Gujarat - 380 009, India}
}

\author{K.~Kong}{
  address={Department of Physics and Astronomy, University of Kansas, Lawrence, KA 66045, USA}
}

\author{C.~Lester}{
  address={Department of Physics, Cavendish Laboratory, JJ Thomson
Avenue, Cambridge, CB3 0HE, UK}
}

\author{K.~Matchev}{
  address={Department of Physics, University of Florida, Gainesville, FL 32611, USA}
}

\author{M.~Park}{
  address={Department of Physics, University of Florida, Gainesville, FL 32611, USA}
}

\begin{abstract}
We revisit the process of transversification and agglomeration of particle momenta that are often performed in analyses at hadron colliders, and 
show that many of the existing mass-measurement variables proposed for hadron colliders 
are far more closely related to each other than is widely appreciated, and
indeed can all be viewed as a common mass bound specialized for a variety of purposes. 
 \end{abstract}

\maketitle

%%%%%%%%%%%%%%%%%%%%%%%%%%%%%%%%%%%%%%%%%%%%
%% MAINMATTER
%%%%%%%%%%%%%%%%%%%%%%%%%%%%%%%%%%%%%%%%%%%%

%%%%%%%%%%%%%%%%%%%%
%\section{Introduction}
%%%%%%%%%%%%%%%%%%%%

%%%%%%%%%%%%%%%%%%%%
\section{Transversification and Agglomeration}
%%%%%%%%%%%%%%%%%%%%
%
%
Almost every analysis of data from hadron colliders uses at some 
point a variable which represents a ``projection'' of an energy or 
momentum into the plane transverse to the beams (see \cite{Barr:2010zj,Barr:2011xt} for recent reviews.). The typical reason for performing 
these projections is that one does not wish the analysis to be 
sensitive to the unknown momentum. 
When it comes to projecting geometric 3-vectors like $\vec{P}$,
the decomposition $\vec{P} \equiv \ourVec{ \vec{p}_\genericT, p_{z} }$ 
is unambiguous. One has no other choice -- the very
definition of the transverse plane requires one simply to dispose of
the $z$-component to arrive at $\vec{p}_{\genericT}=\ourVec{p_x,p_y}$.  
%All the transverse projections considered in this article mu1+st share this property, or else they cannot justify being so named.
However, ``projecting'' the time-like component $E$ is not a well defined operation. 
There is not a single correct answer, but rather a number of different answers, each with
different properties and motivations.  How one should (and even {\em
whether} one should) project time-like components of (1+3) Lorentz
vectors is dependent on what one is trying to achieve.
In the particle physics literature, one can find evidence of at least
three different types of ``transverse projection'' being applied to
(1+3)-Lorentz vectors (summarized in Table \ref{tab:comparisonoftransversemethods}), 
although this diversity is not obvious at first glance, as the majority of papers do not explicitly state which
projection they are using \cite{Barr:2011xt}.
%
%
%\begin{center}
\begin{table}[ht]
\renewcommand\arraystretch{1.5}
\begin{tabular}{|c|c|c|c|}
\hline
               & \multicolumn{3}{c|}{{\bf Transverse projection method} }   \\
\cline{2-4}
{\bf Quantity} & {\bf Mass-preserving `\ourT'} & {\bf Speed-preserving `$\ourperp$'} & {\bf Massless `$\massless$'} \\
\hline
Original (4)-momentum & \multicolumn{3}{c|}{$ P^\mu = \ourVec{ E, \vec{p}_\genericT, p_{z}}$ }  \\ 
(1+3)-mass invariant  & \multicolumn{3}{c|}{$ M = \sqrt{E^2 - \vec{p}_\genericT^{\,2} - p_z^2}$ } \\ 
Transverse momentum & \multicolumn{3}{c|}{ $ \vec{p}_\genericT \equiv \ourVec{p_x,p_y} $ } \\
\hline
(1+2)-vectors & 
$ p_\ourT^\alpha \equiv \ourVec{ e_\ourT, \vec{p}_\ourT }$ & 
$ p_\ourperp^\alpha \equiv \ourVec{ e_\ourperp, \vec{p}_\ourperp }$ &
$ p_\massless^\alpha \equiv \ourVec{ e_\massless, \vec{p}_\massless }$ \\
\hline
\parbox{3.5cm}{{\vskip 2mm}
Transverse momentum \\ under the projection {\vskip 2mm}} 
& $\vec{p}_\ourT \equiv \vec{p}_\genericT$ &  $\vec{p}_\ourperp \equiv \vec{p}_\genericT$ &  $\vec{p}_\massless \equiv \vec{p}_\genericT$ \\
\hline
\parbox{3.5cm}{{\vskip 2mm} 
Transverse energy \\ under the projection {\vskip 2mm}} & 
$e_\ourT \equiv \sqrt{M^2 + \vec{p}_\genericT^{\,2}}$  & 
$ e_\ourperp \equiv E  \left|{\sin \theta}\right| = |\vec{p}_T|/V $ &
$ e_\massless \equiv |\vec{p}_\genericT|$ \\
\hline
\parbox{3.5cm}{{\vskip 2mm}
Transverse mass \\ under the projection {\vskip 2mm}} & $m_\ourT^2 = e_\ourT^2 - \vec{p}_\ourT^{\,2}$ &  
$m_\ourperp ^2 \equiv  e_\ourperp^2 - \vec{p}_\ourperp^{\,2}$ & 
$m_\massless^2 \equiv  e_\massless^2 - \vec{p}_\massless^{\,2} = 0$\\
\hline
\multirow{2}{*}{\parbox{3.5cm}{{\vskip 2mm} Relationship between transverse quantity and its (1+3) analogue {\vskip 2mm}}}
& $m_\ourT = M$ 
&  $m_\ourperp = M \left|\sin\theta\right| $ & $m_\massless = 0$  \\ 
\cline{2-4}
   & \parbox{3.5cm}{{\vskip 2mm} $\frac{1}{v_{\ourT}}=\frac{1}{V}\sqrt{1+(1-V^2)\frac{p_z^2}{p_T^2}}$ {\vskip 1mm}}    & $v_{\ourperp}=V$  &  $v_\massless = 1$  \\ [2mm]
\hline
\parbox{3.5cm}{Equivalence classes under $(1+3) \stackrel{\mathrm{proj}}{\longmapsto} (1+2)$ } &
\parbox{3.5cm}{{\vskip 2mm} 
All $P^\mu$ with the same \\ $p_x$, $p_y$ and  $M$ {\vskip 2mm}} &
\parbox{3.5cm}{{\vskip 2mm} 
All $P^\mu$ with the same \\ $p_x$, $p_y$ and  $V$
{\vskip 2mm}} & 
\parbox{3.5cm}{{\vskip 2mm} 
All $P^\mu$ with the same \\ $p_x$ and $p_y$ 
{\vskip 2mm}}\\ [2mm]
\hline
\end{tabular}
\caption{A comparison of the three transversification methods.}
\label{tab:comparisonoftransversemethods} 
\end{table}
%\end{center}

In forming transverse kinematic variables for composite particles,
one needs to perform two separate operations:
summation of the momentum vectors of the daughter particles,
and projecting into the transverse plane.
The {\em order} of these operations 
does not matter for the two {\em space-like} vector components. 
However, projecting before or after the sum can make a very significant 
difference to the value of the time-like component of the final (1+2) vector ({\it i.e.,} $e_\ourT$, $e_\ourperp$ or $e_\massless$ in Table \ref{tab:comparisonoftransversemethods})  
-- and therefore the operations of 
projecting and summing do {\em not} generally commute:

%%%%%%%%%%%%%%%%%%%%
\section{Interpretation and Generalization}
%%%%%%%%%%%%%%%%%%%%

We define the general procedure that can be used to 
construct the mass-bound variables. 
We describe a broad class of such variables, 
where each individual variable $M_{\{indices\}}$ will be labelled by 
a certain set of indices $\{indices\}$ indicative of the 
way the particular variable was constructed, namely:
\begin{itemize}
\item We are targeting the general event topology, 
where we imagine the inclusive production of $N$ parents, 
each one of our variables will necessarily carry a 
corresponding index $N$. In the process of constructing 
such a variable, we will have to partition (and then agglomerate) the observed
visible particles in the event into $N$ groups ${\cal V}_a$, 
$(a=1,2,\ldots,N)$.
We will then form the (1+3) dimensional invariant mass of each parent 
$\parent a$
\beq
{\cal M}_{a} \equiv \sqrt{g_{\mu\nu}\, (\comp{P}_{a}^\mu+\comp{Q}_{a}^\mu)(\comp{P}^\nu_{a}+\comp{Q}^\nu_a)},
\label{defMa}
\eeq
which is constructed out of the (1+3) momenta $\comp{P}^\mu_{a}$ 
and $\comp{Q}^\mu_{a}$ of the visible and invisible composite daughter particles, respectively.
\item Optionally, instead of the (1+3) dimensional parent mass (\ref{defMa}),
we may choose to consider the corresponding early-partitioned
(late-projected) transverse mass or the late-partitioned (early-projected) transverse mass 
\beq
{\cal M}_{a\genericT}\equiv 
\sqrt{g_{\alpha\beta}\,(\comp{p}_{a\genericT}^\alpha+\comp{q}_{a\genericT}^\alpha)
      (\comp{p}^\beta_{a\genericT}+\comp{q}^\beta_{a\genericT})} , 
~~~~~~~~
{\cal M}_{\genericT a}\equiv 
\sqrt{g_{\alpha\beta}\,(\comp{p}_{\genericT a}^\alpha+\comp{q}_{\genericT a}^\alpha)
                       (\comp{p}^\beta_{\genericT a}+\comp{q}^\beta_{\genericT a})},
\eeq 
where $\comp{p}^\alpha_{a\genericT}$, $\comp{p}^\alpha_{\genericT a}$,
$\comp{q}^\alpha_{a\genericT}$ and $\comp{q}^\alpha_{\genericT a}$
are the (1+2) dimensional momentum vectors for visible and invisible sectors before and after partitioning, 
and the index $T$ takes values in $\ourSet{\ourT,\ourperp,\massless}$, as described in Table \ref{tab:comparisonoftransversemethods}.

\item The last step is to consider the {\em largest} hypothesized parent mass
($\max\ourMaxMinBracs{{\cal M}_{a}}$, $\max\ourMaxMinBracs{{\cal M}_{aT}}$ 
or $\max\ourMaxMinBracs{{\cal M}_{Ta}}$
as appropriate) and {\em minimize} it over all possible values of the 
unknown invisible momenta consistent with the constraints. This minimization 
is always a well-defined, unambiguous operation, 
which yields a unique numerical answer \cite{Konar:2008ei},
\bea
M_N \equiv \min_{\substack{
\sum \vec{q}_{iT} = \mptvec}} 
\ourMaxMinBracs{\max_a\ourMaxMinBracs{{\cal M}_{a}} }, ~~~~~~
M_{N\genericT} \equiv \min_{\substack{
\sum \vec{q}_{iT} = \mptvec}} 
\ourMaxMinBracs{\max_a\ourMaxMinBracs{{\cal M}_{a\genericT}}}, ~~~~~~
M_{\genericT N} \equiv \min_{\substack{
\sum \vec{q}_{iT} = \mptvec}} 
\ourMaxMinBracs{\max_a\ourMaxMinBracs{{\cal M}_{\genericT a}} } \, .
\eea
The minimization over the unknown parameter is performed in order to guarentee that 
the resultant variable cannot be larger than the mass of the heaviest parent,
resulting in an event-by-event lower bound on the mass of the heaviest parent.
\end{itemize}

\begin{figure}
\centerline{
\includegraphics[width=0.41\linewidth]{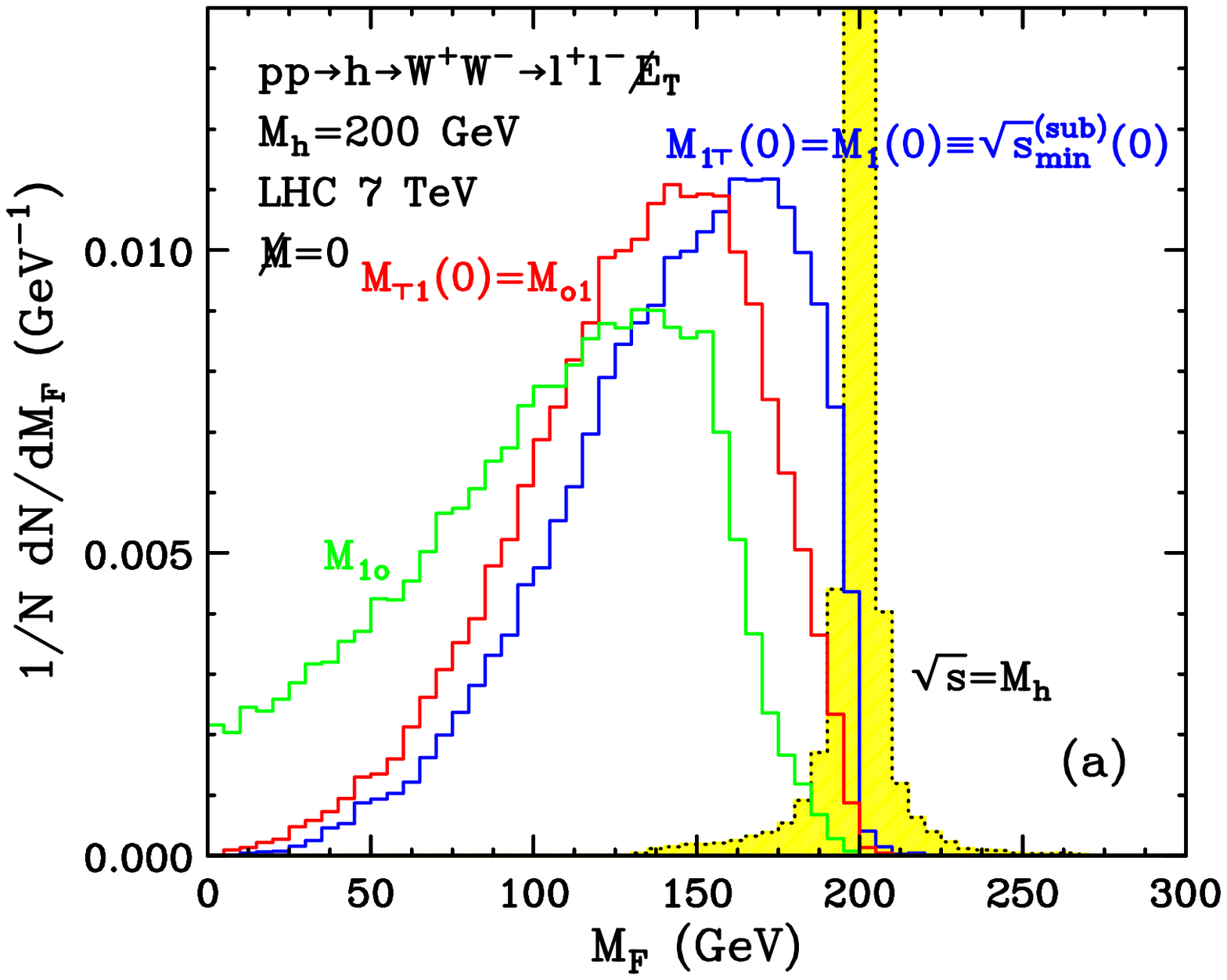} \hspace*{1cm}
\includegraphics[width=0.395\linewidth]{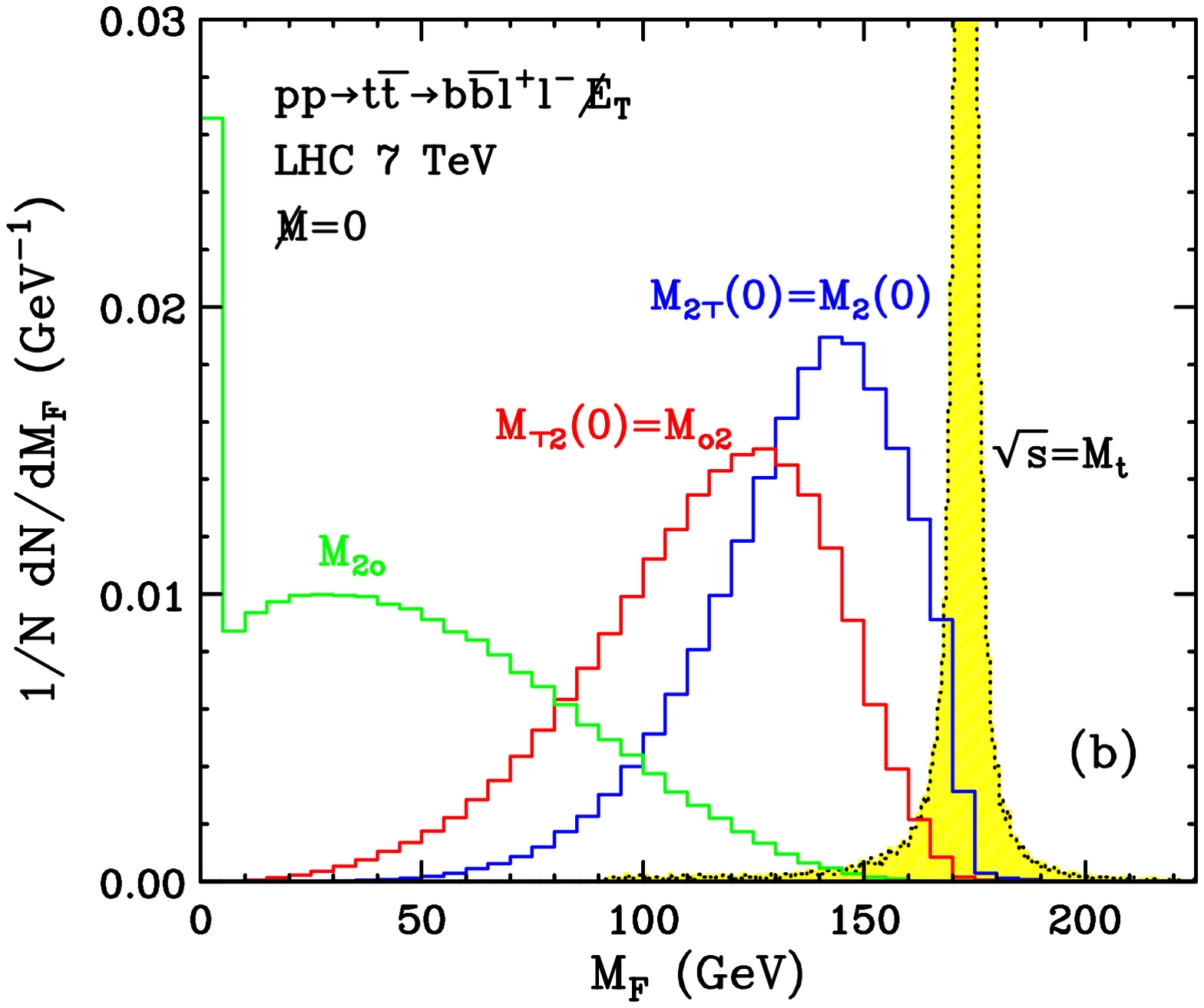} 
 \caption{
(a) Unit-normalized distribution of the five $\numparents=1$ mass-bound variables
$M_{\flox}$, ${\flox}\in \{1, 1\ourT, \ourT 1, 1\massless, \massless 1\}$
for the inclusive Higgs production process
$h\to W^+W^-\to \ell^+\ell^-+\mpt$ at a 7 TeV LHC, with $m_h=200$ GeV
and $\mmass=0$.
The dotted (yellow-shaded) histogram gives the true $\sqrt{\hat{s}}$
distribution, which in this case is given by the Breit-Wigner $h$ resonance. 
(b) The unprojected $M_2$ and the singly projected variables $M_{2\ourT}$, $M_{\ourT2}$, $M_{\massless2}$
and $M_{2\massless}$ for $\numparents=2$ and $t\bar{t}$ example. }
\label{fig:ex}}
\end{figure}

We emphasize that these variables turn out not to be a function of the individual invisible mass hypothesis, but 
instead turn out to be a function of the set, $\chiM=\ourSet{{\mmass_a}\mid{a\in\parentset}}$, 
containing the $\numparents$ ``invisible mass-sum parameters, $\mmass_a$'' defined by
$\mmass_a\equiv \sum_{i\in \invisassign{a}} \tilde M_i$. 
These mass parameters are simple arithmetic sums 
of the hypothesized masses of the individual invisible particles 
associated with any given parent $\parent a$.  
One can also prove from above minimization that the early - agglomerated (late projected) ``'transverse'' 
variables are ``secretly'' (1+3) dimensional, $M_{N\ourT} (\mmass) = M_N (\mmass)$.
This identity reveals that ``transverse'' quantities
do not necessarily ``forget'' about relative longitudinal 
momenta. In particular, this relation teaches us that
whenever the composite particles are formed {\em before}
the transverse projection, the information about 
the relative longitudinal momenta is retained, 
and the result is the same as if everything 
was done in (1+3) dimensions throughout. As a result,
$M_{\numparents\ourT}$ automatically inherits all the advantages 
and disadvantages of its (1+3) cousin $M_{\numparents}$.
These basic variables ( ``unprojected'' $M_N$ and the ``singly projected'' $M_{NT}$ and $M_{TN}$ variables) are shown in Fig.~\ref{fig:ex} for some examples. 
This basic set of variables can be further extended, by considering a second level of projections {\em within} the transverse plane \cite{Barr:2010zj,Konar:2009wn}.

The guiding principle we employ for creating useful hadron-collider event variables,
is that: {\em we should place the best possible bounds on any Lorentz invariants 
of interest, where it is not possible to determine the actual values of those 
Lorentz invariants due to incomplete event information}.  Such incomplete 
information could take the form of lack of knowledge of the longitudinal 
momentum of the primary collision, or lack of knowledge of the 4-momenta 
of individual invisible particles, or lack of knowledge of the number of 
invisible particles which were present, etc.
We contrast this principle with the alternative approach that is used to 
motivate event variables without any explicit regard to whether they have 
an interpretation as an optimal bound of a Lorentz invariant.  
This alternative approach tends to recommend the use of variables 
that are somewhat ad-hoc, but by construction possess useful 
invariances (such as invariance under longitudinal boosts) which are 
designed to remove sensitivity to quantities that are unknown.  
Examples are  $\mptvec$, $h_T$ and $\MEFF$. 
A careful study of similarities and differences of these mass-bound variables 
not only gives insights into why (and under what circumstances) 
these choices are appropriate, it also fits them into a common 
framework -- from which it is straightforward to make 
generalizations to more complex decay topologies. 
Well known kinematic variables such as $M_{T2}$ \cite{Barr:2010zj,Lester:1999tx,Barr:2003rg,Burns:2008va} and $\sqrt{s}_{min}$ \cite{Konar:2008ei,Konar:2010ma} are a special case of the mass-bound variables.

%%%%%%%%%%%%%%%%%%%%%%%%%%%%%%%%%%%%%%%%%%%%%%%%
%% BACKMATTER
%%%%%%%%%%%%%%%%%%%%%%%%%%%%%%%%%%%%%%%%%%%%%%%%

\begin{theacknowledgments}
KK is supported partially by the National Science Foundation under Award No. EPS-0903806 and 
matching funds from the State of Kansas through Kansas Technology Enterprise Corporation.
\end{theacknowledgments}

%%%%%%%%%%%%%%%%%%%%%%%%%%%%%%%%%%%%%%%%%%%%%%%%
%% The bibliography can be prepared using the BibTeX program or
%% manually.
%%
%% The code below assumes that BibTeX is used.  If the bibliography is
%% produced without BibTeX comment out the following lines and see the
%% aipguide.pdf for further information.
%%
%% For your convenience a manually coded example is appended
%% after the \end{document}
%%%%%%%%%%%%%%%%%%%%%%%%%%%%%%%%%%%%%%%%%%%%%%%%

%%%%%%%%%%%%%%%%%%%%%%%%%%%%%%%%%%%%%%%%%%%%%%%%
%% You may have to change the BibTeX style below, depending on your
%% setup or preferences.
%%
%%
%% For The AIP proceedings layouts use either
%%%%%%%%%%%%%%%%%%%%%%%%%%%%%%%%%%%%%%%%%%%%

\bibliographystyle{aipproc}   % if natbib is available
%\bibliographystyle{aipprocl} % if natbib is missing

%%%%%%%%%%%%%%%%%%%%%%%%%%%%%%%%%%%%%%%%%%%
%% You probably want to use your own bibtex database here
%%%%%%%%%%%%%%%%%%%%%%%%%%%%%%%%%%%%%%%%%%%
%\bibliography{sample}

%%%%%%%%%%%%%%%%%%%%%%%%%%%%%%%%%%%%%%%%%%%
%% Just a reminder that you may have to run bibtex
%% All of it up to \end{document} can be removed
%% if you don't like the warning.
%%%%%%%%%%%%%%%%%%%%%%%%%%%%%%%%%%%%%%%%%%%
%\IfFileExists{\jobname.bbl}{}
% {\typeout{}
 % \typeout{******************************************}
 % \typeout{** Please run "bibtex \jobname" to optain}
 % \typeout{** the bibliography and then re-run LaTeX}
 % \typeout{** twice to fix the references!}
%  \typeout{******************************************}
%  \typeout{}
%  }

\end{document}